\documentclass{article}

\usepackage{arxiv}

\usepackage[utf8]{inputenc} 
\usepackage[T1]{fontenc}    
\usepackage{hyperref}       
\usepackage{url}            
\usepackage{booktabs}       
\usepackage{amsfonts}       
\usepackage{nicefrac}       
\usepackage{microtype}      
\usepackage{lipsum}		
\usepackage{graphicx}
\usepackage{natbib}
\usepackage{doi}
\usepackage[ruled,vlined]{algorithm2e}
\usepackage{tikz}
\usepackage{graphicx}
\usepackage{subcaption}
\usepackage{multicol}
\usepackage{soul}
\usepackage{booktabs}

\usepackage{enumitem}
\usepackage{amsmath}
\usepackage{amsthm}
\usepackage{amssymb}
\usepackage{placeins}
\newcommand{\ourproto}{ASHWAchain}
\newcommand{\ourapproach}{Tiramisu}
\newtheorem{proposition}{Proposition}
\newtheorem{theorem}{Theorem}
\newtheorem{Claim}{Claim}
\newtheorem{definition}{Definition}

\title{Tiramisu: Layering Consensus Protocols for Scalable and Secure Blockchains}

\author{ \href{https://orcid.org/0000-0002-4637-4708}{\includegraphics[scale=0.06]{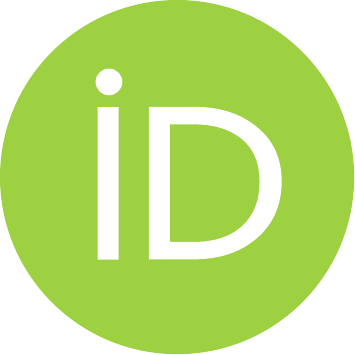}\hspace{1mm}Anurag Jain}\\
	Machine Learning Laboratory\\
	International Institute of Information Technology\\
	Hyderabad, TN, India 500032\\
	\texttt{anurag.jain@research.iiit.ac.in} \\
	\And
	\href{https://orcid.org/0000-0002-5821-8294}{\includegraphics[scale=0.06]{orcid.pdf}\hspace{1mm}Sanidhay Arora} \\
	University of Oregon\\
	Eugene, US \\
	\texttt{sanidhay@uoregon.edu} \\
	\And
	\href{https://orcid.org/0000-0003-1460-6102}{\includegraphics[scale=0.06]{orcid.pdf}\hspace{1mm}Sankarshan Damle} \\
	Machine Learning Laboratory\\
	International Institute of Information Technology\\
	Hyderabad, TN, India 500032\\
	\texttt{sankarshan.damle@research.iiit.ac.in} \\
	\And
	\href{https://orcid.org/0000-0003-4634-7862}{\includegraphics[scale=0.06]{orcid.pdf}\hspace{1mm}Sujit Gujar} \\
	Machine Learning Laboratory\\
	International Institute of Information Technology\\
	Hyderabad, TN, India 500032\\
	\texttt{sujit.gujar@iiit.ac.in} \\
}

\renewcommand{\headeright}{Technical Report}
\renewcommand{\undertitle}{\ \ Technical Report}
\renewcommand{\shorttitle}{Layering Consensus Protocols for Scalable and Secure Blockchains}

\hypersetup{
pdftitle={Tiramisu: Layering Consensus Protocols for Scalable and Secure Blockchains},
pdfsubject={cs.DC},
pdfauthor={Anurag Jain, Sanidhay Arora, Sankarshan Damle, and Sujit Gujar},
pdfkeywords={Blockchain, Scalability},
}

\begin{document}
\maketitle

\begin{abstract}
Cryptocurrencies are poised to revolutionize the modern economy by democratizing commerce. These currencies operate on top of blockchain-based distributed ledgers. Existing permissionless blockchain-based protocols offer unparalleled benefits like decentralization, anonymity, and transparency. However, these protocols suffer in performance which hinders their widespread adoption.  In particular, high time-to-finality and low transaction rates keep them from replacing centralized payment systems such as the Visa network. Permissioned blockchain protocols offer attractive performance guarantees, but they are not considered suitable for deploying decentralized cryptocurrencies due to their centralized nature. Researchers have developed several multi-layered blockchain protocols that combine both permissioned and permissionless blockchain protocols to achieve high performance along with decentralization.
The key idea with existing layered blockchain protocols in literature is to divide blockchain operations into two layers and use different types of blockchain protocols to manage each layer. However, many such works come with the assumptions of honest majority which may not accurately reflect the real world where the participants may be self-interested or rational. These assumptions may render the protocols susceptible to security threats in the real world, as highlighted by the literature focused on exploring game-theoretic attacks on these protocols. We generalize the ``layered'' approach taken by existing protocols in the literature and present a framework to analyze the system in the BAR Model and provide a generalized game-theoretic analysis of such protocols. Using our analysis, we identify the critical system parameters required for a distributed ledger's secure operation in a more realistic setting.
\end{abstract}


\section{Introduction}
\label{sec:intro}


\emph{Bitcoin} \cite{nakomoto} promised to transform the financial system by proposing a decentralized peer-to-peer currency. \emph{Miners} maintain this system by honestly following the underlying consensus protocol through appropriate incentives/rewards. In Bitcoin, and similar cryptocurrencies such as Ethereum, every miner validates transactions and tries to append a set of valid transactions to the set of validated transactions, i.e., append a block on the blockchain. These miners compete against one another in a ``mining'' game. Typically, these miners have to produce ``Proof of Work" (PoW) to write new transaction data to the blockchain.

As of December 2021, Bitcoin's and Ethereum's network processes an average of 4 and 15 transactions per second (TPS), respectively~\cite{blockchaincom_tps, etherscan_tps}. In contrast, Visa's global payment system handles a reported 1,700 TPS and claims to be capable of handling more than 24,000 TPS \cite{visa_claim}. Besides, using such cryptocurrencies for real-time payments, like buying a cup of coffee is not feasible as the protocols only provide \emph{eventual consistency}. It means that each transaction requires a certain number of block confirmations to be confirmed; Bitcoin needs at least 60 mins for confirming a transaction.\footnote{It is also referred to as time to finality.}  


Researchers have proposed several multi-layered protocols that to improve blockchain technology's practical performance for better applicability~\cite{solida, dfninity, peercensus, hybrid-consensus}. These protocols combine both permissioned and permissionless protocols to achieve scalability while maintaining decentralization. We take inspiration from these works and abstract out this layered approach and formalize these layers for different functions of the system. These essentials lead us to a general framework we term as ``Tiramisu''.

We highlight that these works do not provide a game-theoretic analysis to ensure incentive compatibility, i.e., all the above protocols assume miners are either honest or byzantine. However, the miners could also be \emph{strategic players}. That is, they may deviate from the prescribed protocol to gain additional \emph{rewards} because it may not be a strategic miner's \emph{best response} to follow the protocol honestly. Selfish mining, petty mining, and undercutting are some of the strategies that may lead to greater rewards for the miners~\cite{carlsten2016, sapirshtein2016}. In the game-theory literature, a system is said to be \emph{incentive compatible} if it rewards each player greater for playing truthfully as compared to all other possible strategies. We remark that designing such incentive compatible  blockchain protocols that are robust to strategic deviations is a significant challenge. It is also a new area of research with limited prior work~\cite{shoaibaamas}.

\noindent \textbf{Motivation.} Overall, we observe that building a scalable, consistent, and fully decentralized practical blockchain protocol remains elusive. Cryptocurrencies like EOS \cite{eos_whitepaper}, DFINITY \cite{dfninity}, Solida \cite{solida}, etc. have a committee-based blockchain protocol. Among these, PeerCensus \cite{peercensus} and Hybrid Consensus \cite{hybrid-consensus} inherently use a layered approach that demonstrates its potential benefits by achieving higher throughput and faster block confirmations. We highlight none of these works provide a game-theoretic analysis to ensure incentive compatibility and thus, security in the presence of \emph{rational} miners -- the ones maximing their rewards. Hence, the scalability guarantees provided by these protocols might not hold if the protocol is not incentive compatible.

\begin{figure}[tbp]
    \centering
    \includegraphics[height=7cm,width=6cm]{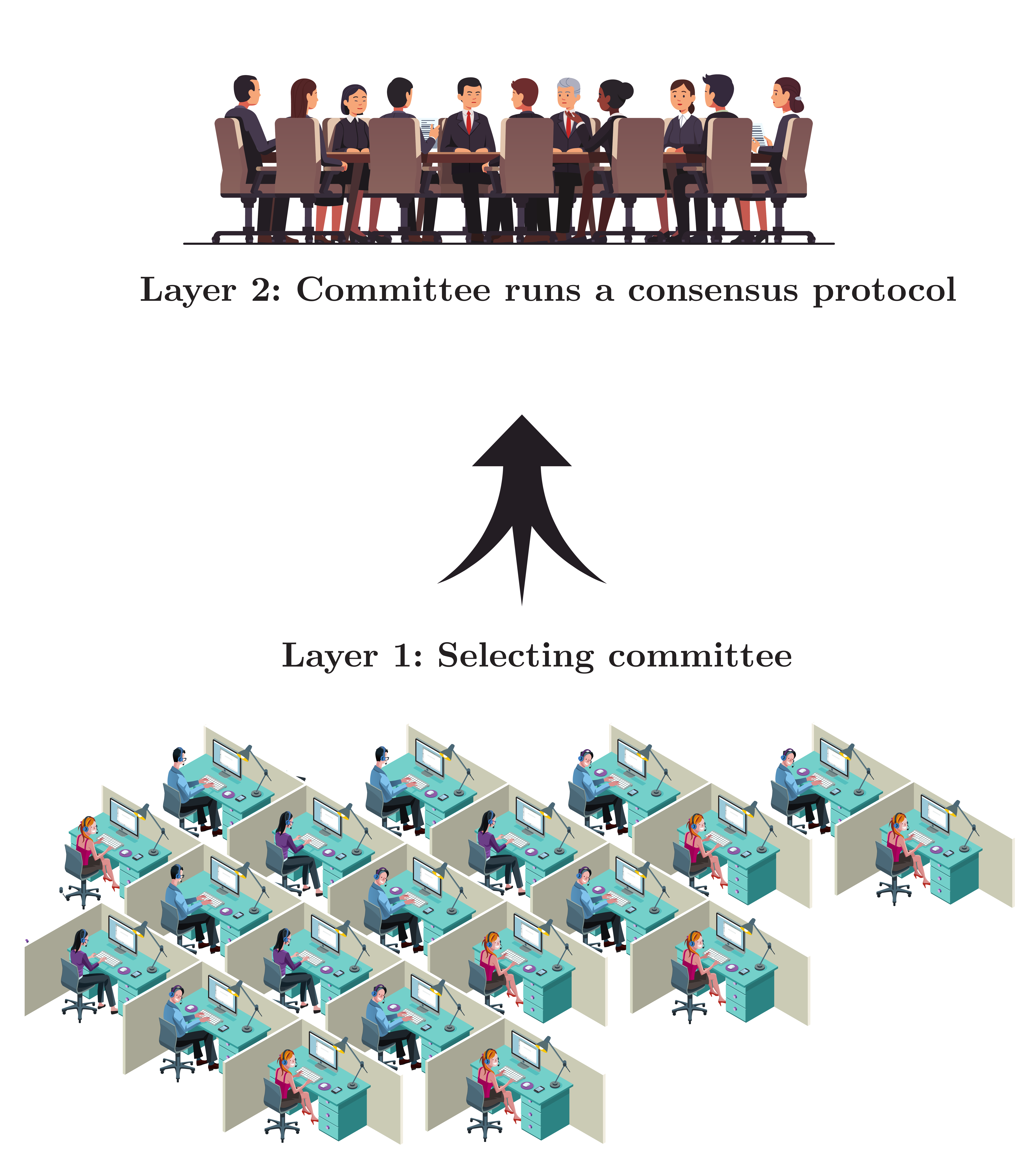}
    \caption{A high-level view of \ourapproach}\label{fig:tiramisu}
\end{figure}

\noindent \textbf{\ourapproach: Overview.} We believe that with a layered-approach, one can create protocols that offer the best of both worlds: throughput and decentralization. We dub this approach as ``\ourapproach'' after the layered dessert. At a high-level, we illustrate \ourapproach\ with Figure~\ref{fig:tiramisu}.

More concretely, \ourapproach\ consists of two layers, namely: Access Control Layer ($ACL$), and Consensus Later ($CSL$). In $ACL
$, we run a permissionless consensus protocol to obtain authorized nodes. Then, with $CSL$, these nodes form a committee to run a BFT consensus protocol on the state of the system. As our framework is general, one can use any permissioned blockchain protocol in CSL, as long as it satisfies certain conditions (refer Section~\ref{sec:tiramisu_analysis}). As standard, our security analysis assumes that $ACL$ is secure. This assumption is based on inherent security guarantees of the underlying consensus protocol that must be carefully employed in the protocol design. Hence, we do not discuss security attacks on it and analyze the security of \ourapproach\ through $CSL$ (Section~\ref{subsec:sa}). We also identify three conditions in our game-theoretic analysis  (Section~\ref{subsec:ga}) that ensure incentive-compatible implementation of any \ourapproach\ protocol. 

\subsection{Contributions.} In summary, our contributions are as follows:


\begin{itemize}
    \item We abstract out the essentials of a layered approach to combine multiple protocols for improved performance of blockchain ecosystem and propose a framework, namely \emph{Tiramisu} (Section~\ref{sec:tiramisu}).


    \item Along similar lines, we present general conditions on protocol parameters, rewards, and the fraction of honest nodes (Section~\ref{subsec:ga}) to ensure an incentive compatible instantiation of a \ourapproach\ framework. Our analysis considers three types of nodes: rational, honest, and Byzantine (Theorem~\ref{thm::PSNE}).

    \item Under the reasonable assumption that the protocol designer uses a secure protocol in ACL, we provide a rigorous security analysis formalism for \ourapproach\ such that the overall system is secure (Section~\ref{subsec:sa}).
    
    \item Analysis with Tiramisu framework also helps in identifying the optimal parameters for a multi-layered blockchain protocol to ensure maximal performance within safe operation.


\end{itemize}

\noindent \textbf{Organisation.} The rest of the paper is structured as follows: First, we discuss and compare our work with the previous related work in Section~\ref{sec:related_work}. Then, in Section~\ref{section:pre}, we outline preliminary concepts for understanding our approach. Next, we propose the \ourapproach\ approach in Section~\ref{sec:tiramisu} and present our analysis of \ourapproach\ in Section~\ref{sec:tiramisu_analysis}. 
Finally, in Section~\ref{sec:conclucion}, we conclude our paper and discuss future work.

\FloatBarrier
\section{Related Work} \label{sec:related_work}

We now discuss the existing literature involving committee-based blockchains, layered approach protocols, and game-theoretic analysis in blockchain protocols, and other scalability solutions.

\noindent \textbf{Committee-based approach.} Committee-based blockchains like EOS \cite{eos_whitepaper} employ byzantine agreement protocols for a scalable blockchain protocol. EOS claims to scale up to 5000 TPS but still has open security concerns \cite{eos_security}.

PeerCensus \cite{peercensus} proposes a protocol that employs a layered approach, which fits as an example in our framework. In PeerCensus, the nodes in the network gain authorized identities by solving a Proof-of-Work puzzle. The nodes controlling these identities then gain privileges to join a \emph{committee} which runs PBFT to produce and validate blocks. We observe that once authorized, these nodes can join and leave this committee during the same session of the protocol. However, we observe that it is not required for a node to leave the committee at any time, and hence these identities are ever-increasing. This observation realizes the following explained drawback for the protocol. The network latency for the exchange of messages in PBFT increases with a message complexity of $O(n^2)$, where $n$ denotes the number of nodes. Now since these nodes are ever-increasing, PeerCensus becomes infeasible to scale because of network latency and time complexity of PBFT limiting the real-world deployment of the protocol. 

Garay et al. \cite{bitcoin-backbone} showed that deferring the resolution of blockchain forks is inefficient and strong consistency provides greater applicability. One might observe that running a committee with only a fraction of the total nodes in the system significantly improves the speed of the protocol due to less number of participants needed for consensus. We use this observation to our advantage when designing our framework.

\noindent \textbf{Layered approach.} Existing consensus protocols most inspire our work in literature that uses a layered approach in some manner. Some of these consensus protocols are used in Algorand \cite{algorand}, OHIE \cite{ohie}, DFINITY \cite{dfninity}, Solida \cite{solida}, PeerCensus \cite{peercensus}, and Hybrid Consensus \cite{hybrid-consensus}. Algorand uses two layers with smart contracts to provide secure and fast common-case transactions and off-chain contracts in layer two for more customization and programming. OHIE composes as many parallel instances of Bitcoin's original backbone protocol as needed to achieve excellent throughput. Dfinity divides its consensus algorithm into four layers tasked to provide identities and registries, random beacon, blockchain, and notarization. Solida is a decentralized blockchain protocol based on a re-configurable Byzantine consensus augmented by proof-of-work. Hybrid Consensus is probably the closest to an instantiation of our approach. It uses a proof-of-stake protocol to manage a fixed-size committee that runs a consensus protocol on transactions.

We take inspiration from these works and aim to build our framework that could capture similar protocols described above. Specifically, we incorporate the notion of dividing tasks into different layers for better abstraction and simplification of both implementation and understanding. The concept of layers also provides independence between different tasks. We believe that this independence may overcome the drawbacks of multiple techniques used in a single layer.

\noindent \textbf{Game-theoretic Considerations.} Solidus \cite{solidus} is the first to consider rational participants by proposing an incentive-compatible Byzantine-Fault-Tolerant protocol for blockchain. However, it does not provide a game-theoretic analysis. Biais et al. \cite{folk} model Bitcoin as a coordination game considering rational participants. Yackolley et al. in \cite{RationalvsByzantine} provide a game-theoretic analysis in consensus-based blockchains considering rational and Byzantine players and model a dynamic game. Manshaei et al. in \cite{shard-based} consider a non-cooperative static game approach for an intra-committee protocol where they show that rational players can free-ride if rewards are equally shared.

Contrary to previous literature, we consider the cost of validating a block in the player's utilities. This consideration makes our work more realistic. We follow the BAR model \cite{BAR} and study rational behavior in a non-cooperative setting, as shown by Halpern et al. in \cite{rational-consensus}. In the BAR model, three types of participants are considered in the system, i.e., honest, rational, and Byzantine.

\noindent \textbf{Scalability Solutions.} Our framework is different from off-chain platforms that use a blockchain like the Bitcoin Lightning network \cite{lightning}, Ethereum's Plasma \cite{plasma}, and Truebit \cite{truebit}. 
Lightning networks and Plasma are decentralized networks using smart contract functionality in the blockchain. Plasma works by creating a tree-like structure of numerous smaller chains. The main idea is to establish a framework of secondary chains that will communicate and interact as sparingly as possible with the main chain. Truebit guarantees correctness in two layers yielding a unanimous consensus layer. In this layer, anyone can object to faulty solutions and an on-chain mechanism that incentivizes participation and ensures fair remuneration. These platforms enhance overall performance and enable instant payments across a network of participants.

These networks are relatively independent and do not always require synchrony with the blockchain, raising additional security and liveness concerns. However, we provide a general framework that splits a blockchain network's operation into its abstracted components and not a synchrony-independent network. Thus, we leverage the advantages from both layers in our framework. 


\section{Preliminaries}\label{section:pre}

In this section, we outline few important preliminary concepts that will be useful for understanding this paper. First, we give a brief of two types of distributed ledgers (DLs). Second, we discuss the properties that distributed ledgers should follow. Next, we discuss BFT protocols. We then discuss permissioned and permissionless protocols. Further, we describe the performance metrics of a given blockchain. And finally, we explain our network model.




\subsection{Permissioned and Permissionless Blockchain Protocols}


        
        


\emph{Permissionless blockchains} are the blockchains that onboard new participants without requiring the permission of existing participants. In this paper, we refer to these participants as \emph{nodes}. Thus, no particular entity can be said to have complete control over the functioning of the system. They are also known as public blockchains. Popular blockchains such as Bitcoin, Ethereum, Litecoin, Dash, and Monero fall under this category. These blockchains allow anyone to transact and join as a validator. The data on these blockchains is publicly available, and complete copies of the ledgers are stored across the network.

\emph{Permissioned blockchains} can be considered as an additional blockchain security system. They maintain an access control layer to allow pre-defined actions to be performed only by specifically identifiable participants. For this reason, these blockchains differ from public and private blockchains.

The main distinction between them is that a permissioned blockchain needs prior approval before use, whereas a permissionless blockchain lets anyone participate in the system. Ethereum~\cite{ethereum}, and Hyperledger~\cite{hyperledger} are widely adopted permissionless and permissioned blockchain protocols, respectively.

\subsection{Properties of Distributed Ledgers}\label{sec:consensus-prop}
We now formally list some important and necessary properties of distributed ledgers. Garay et al.~\cite{BBP} propose important properties of \emph{Liveness} and \emph{Safety} for blockchain protocols to meet to create a secure distributed ledger.

\begin{definition}[Liveness~\cite{BBP}]
\emph{Liveness} with parameters $u, \kappa$ states that if all parties attempt to include a valid transaction in the ledger at time $t$, then the transaction would be accepted by every honest party with security parameter $\kappa$ at a time $t+u$ with high probability.
\end{definition}

\begin{definition}[Safety~\cite{BBP}]
\emph{Safety} with parameter $\kappa$ states that if an honest party has accepted a transaction with a security parameter $\kappa$ at time $t$, then at any time after $t+\delta$ the transaction must be present in the longest chain available with every honest party.
\end{definition}

\noindent In contrast, consensus requires the following properties:
\begin{itemize}[leftmargin=*]
    \item \emph{Agreement} - All non-faulty nodes decide on the same value.
    \item \emph{Termination} - All non-faulty nodes terminate in finite time.
    \item \emph{Validity} - If all non-faulty nodes propose the same value, then no other value can be decided.
\end{itemize}

We now describe consensus protocols that also incorporate byzantine fault tolerance.

\subsection{Byzantine Agreement Protocols} \label{sec:bft}

A Byzantine fault is a condition of a distributed system, where components (or nodes) may fail, and there is imperfect information on whether a component has failed or not. Byzantine Fault Tolerance (BFT), derived from Byzantine Generals’ Problem, is the property of a distributed network to reach consensus even with some amount of byzantine fault in the system. The objective of a BFT mechanism is to safeguard against system failures. Byzantine agreement protocols or BFT protocols strive to provide the property of Byzantine Fault Tolerance.

The most widely known and used BFT agreement protocol is the Practical Byzantine Fault Tolerance~\cite{pbft}, commonly known as PBFT. One might observe that these protocols can reach consensus with absolute finality fairly quickly if there are fewer participants. However, their speed reduces drastically with every increasing node in the system.



\subsection{Network Model}
Here, we formally describe our network model. We consider a partially synchronous peer-to-peer network consisting of participants who control identities in the network. These identities are denoted by their public-private ($pk, sk$) key pair and hence, are only pseudonyms since they do not leak any real-world information about the participant\footnote{For brevity, we use identities in place of pseudonymous identities.}. Participants connect by a broadcast network over which they can send messages to everyone.

\subsubsection{Agent Model\label{bar_model}}
We follow the BAR Model introduced by~\cite{BAR} that consists of Byzantine, Altruistic (or Honest) and Rational Parties.
\begin{definition}[$\mathcal{H}$ Honest Party]
A party is said to be \emph{honest} if and only if it strictly follows the protocol.
\end{definition}
\begin{definition}[$\mathcal{R}$ Rational Party]
A party is said to be \emph{rational} or \emph{self-interested} if it may strategically deviate from the protocol if the deviation is expected to yield a higher reward.
\end{definition}
\begin{definition}[$\mathcal{B}$ Byzantine Party]
A party is said to be an \emph{adversarial} or \emph{byzantine} if it may or may not follow the protocol. The adversary's goal is to disrupt the operation of the protocol, and it does \textbf{not} try to maximize its reward.
\end{definition}



With this background, we now present \ourapproach\ in the next section.

\begin{table}[t]
    \centering
    \begin{tabular}{cc}
    \toprule
         \textbf{Notation} & \textbf{Meaning}\\
         \midrule
         $\alpha_{participant}$ & Fraction of network resources owned by a participant\\
         $\alpha_A$ & Fraction of network resources owned by adversary $A$\\
         $\Pi_{ACL}$ & Underlying Access Control Layer Protocol\\
         $\Pi_{CSL}$ & Underlying Consensus Layer Protocol\\
         $\Pi_{\mathcal{TM}}$ & \ourapproach\ Protocol: $\{\Pi_{ACL}, \Pi_{CSL}\}$ \\
         $n_{CSL}$ & Total number of committee nodes in $CSL$.\\
         $n_f$ & Minimum nodes needed to compromise $\Pi_{CSL}$.\\
    \bottomrule
    \end{tabular}
    \caption{Protocol specific notations}
    \label{tab:pre-notations}
\end{table}

\begin{table}[t]
\centering
\begin{tabular}{c c} 
   \toprule
    \textbf{Term} & \textbf{Definition}\\
    \midrule
    $\mathcal{N}$ & Set of nodes, $\{1, 2, \dots, n_{CSL}\}$\\
    $\tau_i$ & Node $i$'s type, i.e.,$\tau_i\in\{\mathcal{H},\mathcal{R},\mathcal{B}\}$\\
    $s^1_i$ & Node $i$ signs the block without validating 
    \\
    $s^2_i$ & Node $i$ signs the block only if its valid\\
    $s^3_i$ & Node $i$ signs and proposes only invalid blocks\\
    $S_i$ & Set of pure strategies of nodes $i$, $\{s^1_i, s^2_i, s^3_i\}$\\
    $TR_i$ & Total rewards of node $i$ for a transaction block.\\
    $u_i$ & Utility of node $i$; $u_i: \tau_i \times TR_i \times S \rightarrow \mathbb{R}$\\
\bottomrule
\end{tabular}
\caption{Game constituents} \label{game-table}
\end{table}

\begin{table}[t]
\centering
\begin{tabular}{c p{6.5cm}}
\toprule
    \textbf{Term} & \textbf{Definition}\\
    \midrule
    $c_{mine}$ & The cost of mining a block in $ACL$.\\ 
    $c_{val}$ & The cost to validate one byte of data.\\ 
    $n_{TX}$ & Number of transaction blocks for a given committee session.\\
    $\kappa_i$ & Cost incurred by a rational node $i$ if an invalid transaction block is accepted.\\ 
    $\kappa_\mathcal{R}$ & $\kappa_i$ for the rational node with minimum stake invested.\\ 
    $\phi$ & Number of bytes of shared state to be validated for a single block.\\ 
    $TR$ & Minimum $TR_i$ for the transaction block.\\
    $P_{invalid}$ & Belief probability of invalid block being accepted, $P_{invalid}: \mathcal{N}^{n_{CSL}} \times S_i \rightarrow  [0, 1]$.\\
    \bottomrule
\end{tabular}
\caption{Relevant Game Notations}     \label{notations-table}
\end{table}

\begin{table}


\centering
    \begin{tabular}{cc} 
        \toprule
        \textbf{Name} & \textbf{State}\\
        \midrule
        \textsc{PowChain} & Proof of Work Blockchain\\ 
        $T$ & List of transaction validators\\ 
        \textsc{ComChain} & Committee Blockchain\\ 
        \textsc{TxChain} & Transaction Blockchain\\ 
        $O$ & Operation Log\\ 
        $B$ & Account Balances\\
        \bottomrule
    \end{tabular} 
\caption{Shared state constituents}\label{tab::SS-1}

\centering
    \begin{tabular}{ cc }
        \toprule
        \textbf{Name} & \textbf{Operation}\\
        \midrule
        \texttt{powBlock}$(b)$ & Adds PoW block $b$ to \textsc{PowChain}\\ 
        \texttt{txBlock}$(t)$ & Adds transaction block $t$ to $T$ \\ 
        \texttt{comBlock}$(c)$ & Adds committee block $c$ to \textsc{ComChain} \\
        \bottomrule
    \end{tabular}
\caption{Shared state operations}\label{tab:SS-2}
\end{table}

\section{Tiramisu: A Layered Approach} \label{sec:tiramisu}
In \ourapproach, we split the operation of the blockchain into two independent layers. 
The first and the second layers are called Access Control Layer ($ACL$) and Consensus Layer ($CSL$) respectively. $ACL$ manages nodes in the network via a permissionless consensus protocol. Whereas, $CSL$ employs a permissioned consensus protocol among the authorized nodes from the first layer. These nodes will run a Byzantine agreement protocol to verify transactions and reach a consensus on the state of the system.

Let $\Pi_{\mathcal{TM}} = \{ \Pi_{ACL}, \Pi_{CSL} \}$ be a Tiramisu protocol where $\Pi_{ACL}$ and $\Pi_{CSL}$ refer to the underlying consensus protocols running in the Access Control Layer and the Consensus Layer respectively.

\subsection{Access Control Layer ($ACL$)}
$ACL$ is responsible for providing sybil-resistant node identities to the Consensus Layer.
The identity of these \emph{nodes} are simply a derivative of the public key of a public-secret-key pair 
$\langle pk, sk\rangle$, owned by a participant. Each node is identified by its public address.

In this layer, participants maintain a separate blockchain just to obtain sybil-resistant identities. Participants in this layer run any permissionless blockchain consensus protocol, by choice of protocol design, denoted by $\Pi_{ACL}$. Note that $\Pi_{ACL}$ must satisfy some pre-defined conditions. Towards stating these pre-defined conditions, we first describe a \emph{democratized resource}. A democratized resource is essential to run the protocol $\Pi_{ACL}$. It can be captured as a real-world resource and must be directly responsible for: first, the basis of sybil-resistance in $\Pi_{ACL}$; and second, for obtaining control over specific network resources. These network resources must be directly proportional to the fraction of control that they gain over the protocol $\Pi_{ACL}$. Typically $\Pi_{ACL}$ can be a proof-of-X base protocol like Proof-of-Work. We assume that $ACL$ is secure based on the inherent security of $\Pi_{ACL}$.



\noindent Formally, $\Pi_{ACL}$ must satisfy the following properties:
\begin{itemize}[leftmargin=*]
    \item There must exist a protocol parameter that is directly able to express the democratized resource present in the network.
    \item The probability of a node adding a new block to its blockchain must be linearly dependent on the democratized resource.
\end{itemize}

For instance, $\Pi_{ACL}$ could be a Proof-of-Work protocol where the democratized resource is \emph{computational power} and the protocol parameter could be the \emph{hash-rate} of the network. Another $\Pi_{ACL}$ could be a Proof-of-Stake protocol where both the resource and this parameter are the stake invested in the system itself. 

\noindent \textbf{Operation.} $\Pi_{ACL}$ is run to determine the identities of the nodes that will be participating in the Consensus Layer. Each block in the blockchain of this layer must contain a single public address representing the identity of the node that wishes to join the $CSL$. Once a block reaches finality (Section~\ref{section:pre}), $ACL$ uses a pre-defined interface to interact with $CSL$ and propose the identity present in this block to join the committee in $CSL$.
Observe that any participant will only invest their network resources for the identities of the nodes that they wish to get promoted in the Consensus Layer. The key insight behind the sybil-resistant nodes is that the democratized resources are hard to obtain and may not be scaled at will. Observe that the probability of a node joining the $CSL$ is directly proportional to the amount of democratized resources owned by the participant, denoted by $\alpha_{participant}$. This is a key observation that will be used in our security analysis (Section~\ref{sec:tiramisu_analysis}).

In Section~\mbox{\ref{sec:tiramisu_analysis}}, we show that for the system to function correctly, the fraction of resources controlled by the adversary must be less than $\frac{n_f}{n_{CSL}}$. Here, $n_{CSL}$ and $n_f$ are the total number of nodes in the Consensus Layer
and the minimum number of nodes needed to compromise $\Pi_{CSL}$, respectively. Then we show that $\alpha_A$ is a reasonable bound for practical systems. We now move on to the second layer called the Consensus Layer ($CSL$) in which the committee of nodes from this layer handles the transactions.

\subsection{Consensus Layer ($CSL$)}
This layer is responsible for handling transactions and reaching a consensus on the state of the system. The nodes in this layer are controlled by participants. This layer maintains a shared state of the system by running any Byzantine agreement protocol or BFT protocol, denoted by $\Pi_{CSL}$. One can use any BFT protocol suited for the permissioned blockchain setting as long as it satisfies the network model. The shared-state can only be modified by pre-determined operations. This shared-state can consist of anything related to the purpose of the protocol like running a cryptocurrency, notarization platform, smart-contract platform, etc. For a cryptocurrency, this state may contain account balances, a transaction blockchain, etc. Note that however, for the sake of simplicity in this paper, we use the term \emph{transaction blockchain} to denote the shared state which represents the purpose of the \ourapproach~protocol, $\Pi_{\mathcal{TM}}$. Similarly, we use the term \emph{transaction} throughout this paper to denote a state change in the transaction blockchain.

We now claim that the \ourapproach\ protocol satisfies the properties of distributed ledgers described in Section~\ref{sec:consensus-prop}. BFT protocols structure their execution into a sequence of views, each with a designated leader process. These protocols guarantee Safety and liveness by ensuring that all correct nodes eventually overlap in a single view, with the right leader, for enough time to reach consensus.

\begin{Claim}\label{claim:CSL}
A \ourapproach\ protocol session achieves Safety and liveliness if $\Pi_{CSL}$ does so.
\end{Claim}

At any instance in time, this layer is operated by $n_{CSL}$ nodes, collectively called the \emph{committee} of this layer. The committee size can be either fixed or variable depending on network parameters. Fixed-size can be achieved through various techniques. For example, by allowing only the latest $n_{CSL}$ nodes from the $ACL$ blockchain to join the system. Note that, the shared-state must store a copy of the $ACL$ blockchain, or an equivalent, to ensure verification of the nodes. The nodes in the committee are decided only after a BFT agreement on the identity of these nodes. This agreement ensures that any behavior of the $ACL$ blockchain, like a blockchain fork, is not relevant in this layer. Since we use  Byzantine  agreement  protocols, $CSL$ inherently satisfies the  properties of Safety and liveness. This justifies Claim~\ref{claim:CSL}.

Now we discuss our analysis of \ourapproach, in which we also identify three conditions on protocol parameters that ensure incentive compatible implementation of a \ourapproach\ protocol.



\section{Tiramisu: Analysis} \label{sec:tiramisu_analysis}

It is important to note that each node is independently selected to be on the committee. The probability of this selection is proportional to the number of democratic resources, as required for participating in $ACL$ which is owned by the node in the network.




The BAR model (described in Section~\ref{bar_model}) allows us to analyze the game-theoretic and security aspects of \ourapproach\ as presented next.


\subsection{Game-theoretic Analysis} \label{subsec:ga}

As stated in our player model, the reward structure of any blockchain consensus protocol induces a game among the participants. 
We highlight that we account for the computational costs of validating the state of the system. This consideration forms the basis of our game model, which also makes it more suited for realistic setting.
To analyze this induced game, we define the following notations: Consider a set of nodes $\mathcal{N}=\{1,\dots,n_{CSL}\}$. This set includes all the information about a node, including its strategy and type. Let each participant $i$'s strategy be $s_i$ and its type be $\tau_i$, where $\tau_i \in \{\mathcal{H}, \mathcal{R}, \mathcal{B}\}$. Note that, the strategy will depend on the participant's type. We have $\vec{s}=(s_1,\dots,s_m)$ as the strategy vector and $\vec{s}_{-i}$ as the vector without node $i$. Let $TR_i$ denote participant $i$'s reward for one transaction block. Note that this reward is left for design of the protocol. With this, $u_i(\tau_i, TR_i, \vec{s})$ represents a participant $i$'s utility from its participation. We assume that rational participants do not collude with each other based on the premise that the distributed environment will make it resistant to collusion.

In Tiramisu, we game-theoretically analyze the node's behavior at equilibrium using the following notion.

\begin{definition}[Pure Strategy Nash Equilibrium (PSNE)]
    A strategy vector  $\vec{s^*}=\{s_1^*,\ldots,s_m^*\}$ is said to be a Pure Strategy Nash Equilibrium (PSNE) if for every node $i$, it maximizes its utility $u_i(\vec{s}^*,\tau_i,r_i)$ i.e., $\forall i \in \mathcal{M},~\forall j$,
    \begin{equation} \label{equation:PSNE}
    u_i(\vec{s^*},\tau_i,r_i) \geq u_i(s_i,\mathcal{S}_{-i}^*,\tau_i,r_i);~\forall s_i.
    \end{equation}
\end{definition}


Intuitively, PSNE states that it is the best response for a node to follow $\vec{s^*}$, given that every other node is following it.

\noindent \textbf{Equilibrium.} Towards this, we assume that a rational player $\mathcal{R}$ incurs a cost $\kappa_\mathcal{R}$ if any malicious transaction block is committed in $CSL$, similar to \cite{RationalvsByzantine}.
This assumption is based on the premise that the entire ecosystem of the currency inflicts harm when an invalid block is accepted. 
This cost is directly proportional to the amount of stake that the rational node has invested in the system. Note that this stake can be in the form of electricity for mining in a proof-of-work blockchain, amount locked in the blockchain for participation in a proof-of-stake blockchain, rewards earned by the miner, and likewise. For simplicity, we take this proportionality constant to be 1, making the cost equal to the stake invested. 
We denote this cost of a rational node $i$ by $\kappa_i$. Let $\kappa_{\mathcal{R}}$ be $\kappa_i$ of the rational player with minimum invested stake. Observe that the protocol designer can set a lower-bound on $\kappa_{\mathcal{R}}$ using a variety of well-known techniques.

Although Byzantine nodes behave arbitrarily by definition, we consider a specific behavior of Byzantine nodes. Specifically, we assume that the objective of Byzantine nodes is to minimize the utility of the rational nodes and prevent the protocol from achieving its goal, regardless of the cost they incur. Note that this assumption ensures the security of a \ourapproach\  protocol in the worst possible case, and hence it ensures security for all cases, i.e. regardless of the behavior of Byzantine nodes. With these assumptions, we model a game between honest, rational and Byzantine nodes' nodes in the committee, defined as: $\Gamma = \langle N, (\tau_i), (S_i), (u_i) \rangle $.

\begin{definition}[Nash Incentive Compatible] \label{def:ics}
We say that a \ourapproach\ protocol $\Pi_{\mathcal{TM}}$ is Nash Incentive Compatible (NIC) if $\vec{s^*} = (s_1^2, \cdots, s_{n_{\mathcal{R}}}^2)$ is a PSNE for all rational nodes and at this PSNE all rational nodes obtain non-negative utility.
\end{definition}

For the analysis, let $P_{invalid}(\mathcal{N}, s^*_i)$ be node $i$'s belief that an invalid block is accepted after it follows $s^*_i$. Honest nodes will always follow the protocol, i.e. $s^2_i$ strategy. Byzantine nodes will always follow $s^3_i$ because this is the strategy that best aligns with our assumption on the objective of Byzantine nodes. Hence we only consider rational nodes' behavior and therefore their equilibrium strategies.

The expected utility of a rational node, when following $s_{i}^{1}$ strategy for a single transaction block in $CSL$, is total rewards obtained for that block, minus the average cost of mining a block in $ACL$ for that committee session, and minus the expected $\kappa_i$. Similarly, to calculate the utility of a rational node for strategy $s^2_i$, we need to subtract the cost of validating a block as well. The expected utilities of a rational node $i$ for one round are as follows:

\begin{align}\label{utility:s1}
   u_i(\cdot, \cdot, s^1_i) & = TR_i -  \frac{c_{mine}}{n_{TX}} - P_{invalid}(\mathcal{N}, s^1_i) \cdot \kappa_i \\
 u_i(\cdot, \cdot, s^2_i) &  = TR_i - \phi \cdot c_{val} - \frac{c_{mine}}{n_{TX}} - P_{invalid}(\mathcal{N}, s^2_i) \cdot \kappa_i\label{utility:s2}
\end{align}

For $n_{CSL}$ nodes, let $n_{H}$, $n_{R}$, and $n_{B}$ be number of honest, rational and Byzantine nodes respectively. Trivially, $n_{CSL} = n_H + n_R + n_B$. We use PSNE for analysis, as we have a static game with asymmetric information.

\begin{Claim}\label{claim::prob}
For every rational node $i$, the probability of an invalid block being accepted will be more if it signs a block without validating, as compared to when it validates and then signs, i.e., \[\delta = P_{invalid}(\mathcal{N},s^1_i) - P_{invalid}(\mathcal{N},s^2_i) > 0.\]
\end{Claim}

Intuitively, Claim~\ref{claim::prob} follows from the fact that the chance of an invalid block being accepted is more if a rational node decides to sign the block as valid without verifying it. Now we present the formal proof for the claim.

\begin{proof}
Without loss of generality, for the proof we consider a rational player $i$. Let
\begin{equation*}
P_{invalid}(\mathcal{N}, s^1_i) = k_1
\end{equation*}
and
\begin{equation*}
P_{invalid}(\mathcal{N}, s^2_i) = k_2
\end{equation*}
s.t. 
\begin{equation*}
0 < k_1, k_2 < 1.
\end{equation*}

In the event when node $i$ plays the strategy $s^1_i$ i.e. signs the block as valid without validating, its belief regarding an invalid block being accepted can only increase. This follows by observing that the size of the committee, i.e., $n_{CSL}$ is finite. Thus, node $i$ not validating a block will directly imply that the chance of an invalid block being accepted will be more, i.e., $k_1 > k_2$. Now, 
$$
P_{invalid}(\mathcal{N}, s^1_i) - P_{invalid}(\mathcal{N}, s^2_i) = k_1 - k_2 = \delta > 0
$$
\end{proof}

We denote the minimum possible value of $\delta$ for a protocol session running with specific security guarantees as $\delta_{min}$. Along with the node's belief and some protocol parameters, $\delta_{min}$ is dependent on the CDF $F$ which is appropriately defined in security analysis under proof of Proposition~\ref{proposition1}.

\noindent \textbf{NIC Conditions.} For all rational nodes, there exists a PSNE under certain conditions. We identify three such conditions to ensure that a \ourapproach\ protocol session $\Pi_{\mathcal{TM}}$ is NIC (Definition~\ref{def:ics}). We call these set of conditions as \emph{NIC-Conditions} which are based on the following requirements. First, to ensure that $\Pi_{CSL}$ is secure. Second, to ensure that rational nodes attain non-negative utility when they deviate from following the honest strategy. And third, to ensure that each rational node must obtain positive utility. The respective NIC-Conditions are as follows:
\begin{enumerate}
    \item \textbf{Faithful Fault Tolerance Condition} $n_B < n_f$
    \item \textbf{Maximum Payload Condition} $\phi \le \frac{\kappa_{\mathcal{R}} \cdot \delta_{min}}{c_{val}}$
    \item \textbf{Minimum Reward Condition} $TR \ge \phi \cdot c_{val} + \frac{c_{mine}}{n_{TX}}$
\end{enumerate}

Note that NIC-Conditions provide constrains on protocol parameters which are dependent on protocol design, implementation, and live-network data\footnote{Here, live-network data at least includes the stored data of a protocol session along with its consolidated statistics.}. Hence, satisfying these conditions rely on the careful implementation by protocol designer.


\begin{theorem}\label{thm::PSNE}
In Tiramisu, Faithful Fault Tolerance Condition, Maximum Payload Condition and Minimum Reward Condition are sufficient to ensure that the protocol is Nash Incentive Compatible (NIC). Formally, if Faithful Fault Tolerance Condition, Maximum Payload Condition and Minimum Reward Condition are true, then $\Pi_{\mathcal{TM}}$ is NIC according to Definition~\ref{def:ics}.
\end{theorem}

\begin{proof}
To ensure that $\Pi_{\mathcal{TM}}$ is NIC, first requirement is that $\Pi_{CSL}$ must work correctly. Hence the condition $n_B < n_f$ should be true, justifying the requirement of Faithful Fault Tolerance Condition.

Second, $s^*_i = s^2_i$ must be a PSNE for every rational node $i$. To ensure this, $u_i(R, s^1_i) \leq u_i(R, s^2_i)$ should satisfy according to Equation \ref{equation:PSNE}. On solving this condition using equations \ref{utility:s1} and \ref{utility:s2} we get,
\begin{equation*}
    \kappa_i \cdot (P_{invalid}(\mathcal{N}, s^1_i) - P_{invalid}(\mathcal{N}, s^2_i)) \geq \phi \cdot c_{val}.
\end{equation*}

\noindent Substituting ($P_{invalid}(\mathcal{N}, s^1_i) - P_{invalid}(\mathcal{N}, s^2_i)$) as $\delta$, where $\delta \in (0, 1]$ following from Claim \ref{claim::prob}, we get our result for the value of $\phi$ as:
\begin{equation*}
    \phi \cdot c_{val} \leq \kappa_i \cdot \delta.
\end{equation*}
Now, since $\kappa_{\mathcal{R}} \le \kappa_i$ and $\delta_{min} \le \delta$, we get the following condition as:
\begin{equation*}
    \phi \cdot c_{val} \leq \kappa_{\mathcal{R}} \cdot \delta_{min},
\end{equation*}
justifying the requirement of Maximum Payload Condition.

Finally in third, for the above mentioned PSNE, utilities gained by all rational nodes must be non-negative, i.e. 
\begin{equation*}
    TR_i - \phi \cdot c_{val} - \frac{c_{mine}}{n_{TX}} - P_{invalid}(\mathcal{N}, s^2_i) \cdot \kappa_i \ge 0.
\end{equation*}
Notice that in this PSNE, since all rational nodes will validate the blocks, $P_{invalid}(\mathcal{N}, s^2_i) = 0$. Also, $TR_i \le TR$. From this we get the condition as:
\begin{equation*}
    TR \ge \phi \cdot c_{val} + \frac{c_{mine}}{n_{TX}},
\end{equation*}
justifying the Minimum Reward Condition.
\end{proof}

However, in a special case where $n_{H} > n_{CSL} - n_f$, $S^* = (s^1_i)$ is the PSNE for rational nodes where only valid blocks are accepted and is as special case of a NIC $\Pi_{\mathcal{TM}}$.

\begin{figure*}[ht]
    \centering
    \minipage{0.49\linewidth}
        \includegraphics[width=0.9\linewidth, height=5cm]{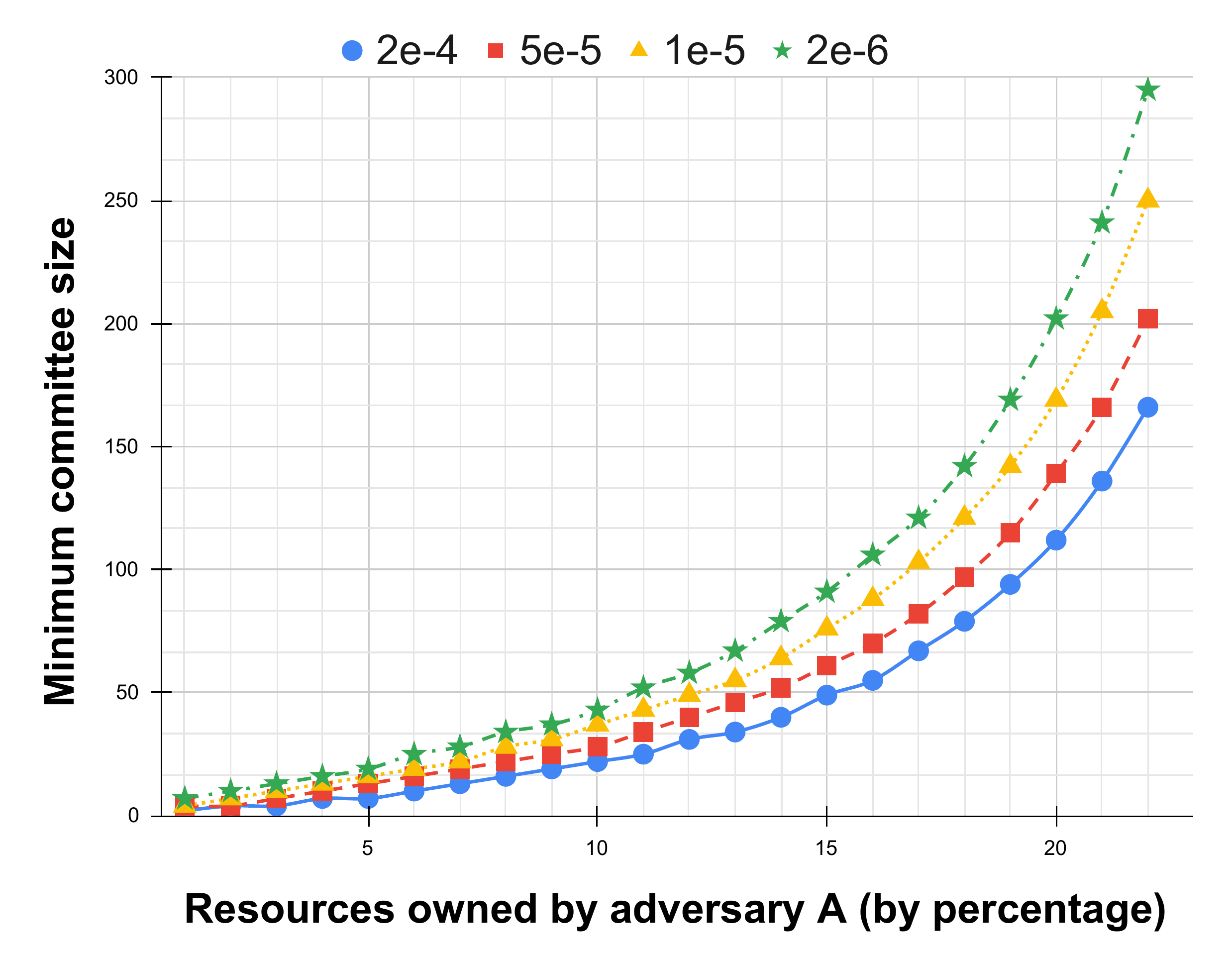}
    \caption{For a given $\alpha_A$, $n_{CSL}^{min}$ for different values of $\epsilon$.}
        \label{fig:spvspop}
    \endminipage\hfill
    \minipage{0.49\linewidth}
        \includegraphics[width=0.9\linewidth]{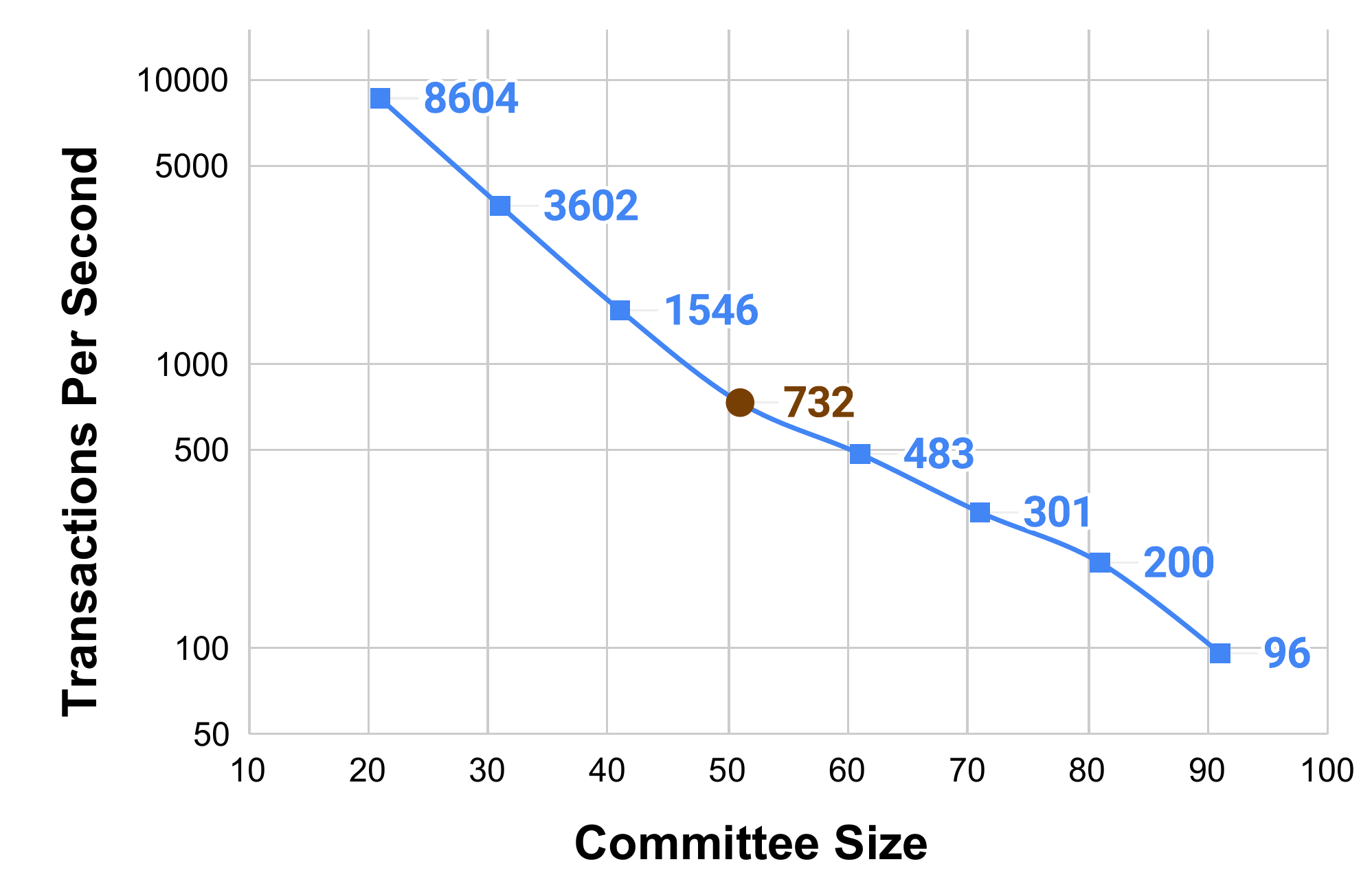}
    \caption{Transactions Per Seconds vs. Committee Size}\label{fig:tps}
    \endminipage\hfill
\end{figure*}
\subsection{Security Analysis}\label{subsec:sa}

\noindent \textbf{Setting.} Our security analysis assumes that protocol $\Pi_{ACL}$ is secure, i.e. $ACL$ runs securely based on inherent security of $\Pi_{ACL}$ and intelligent protocol design. We thus analyse the security of a $\Pi_{\mathcal{TM}}$ session by just ensuring the security of $CSL$.
Notice that \ourapproach's security relies on securing $\Pi_{CSL}$ and the inherent security guarantees of $\Pi_{ACL}$. We achieve this security just by quantifying the committee size in $CSL$ based on an appropriately defined function in this analysis. Next, we consider this security analysis for a NIC \ourapproach\ protocol session. This is followed from Theorem~\ref{thm::PSNE} in our game-theoretic analysis. Notice that in this setting, since all rational players will follow the honest strategy, it is justified to consider all rational nodes as honest. Hence, we consider only honest and Byzantine nodes in our security analysis of a NIC \ourapproach\ protocol. Towards this, we first define our adversary model.

\noindent \textbf{Adversary Model.}
We  consider  an  adversary A controlling $\alpha_A$ fraction of the democratized resources in the network. The remaining  resources  are  controlled  by  a  meta  entity $\mathcal{H}$, denoting all other participants in the network. The goal of the adversary is to control more than $n_{f}$ nodes, i.e., the  minimum number of nodes needed to compromise the protocol.

Note that the probability of a node joining the $CSL$ is directly proportional to the amount of democratized resources owned by the respective participant in $ACL$. We denote the total participants in $CSL$ to be $n_{CSL}$ at any time. Similar to \cite{peercensus}, we analyze Tiramisu in its \emph{steady state}, i.e., the number of validators and computational resources are governed by their respective expected value. To capture this formally, we define the following \emph{secure state}.

\begin{definition}[Secure State]\label{def:SS}
A $\Pi_{\mathcal{TM}}$ session is said to be in a secure state ($\mathcal{SS}$) if the number of nodes in $CSL$ controlled by an adversary is strictly less than minimum number of nodes needed to compromise $\Pi_{CSL}$. Formally, $n_A < n_f$.
\end{definition}

We highlight that Claim~\ref{claim:CSL} relies on security and correct execution of $\Pi_{\mathcal{TM}}$. Since it relies on security guarantees of underlying protocols $\Pi_{ACL}$ and $\Pi_{CSL}$, we say that Claim~\ref{claim:CSL} is true if $\Pi_{\mathcal{TM}}$ session is in a secure state. Hence, we state that a $\Pi_{\mathcal{TM}}$ session achieves Safety and liveness only in a secure state given $\Pi_{CSL}$ does so. Towards this, we introduce the secure state function $F_{SS}$ which provides $n_{CSL}^{min}$, the minimum number of nodes required in $CSL$ for a $\mathcal{SS}$ with $1 - \epsilon$ probability.
$$
F_{SS} : R \times R \rightarrow Z^+,
$$
and
$$
n_{CSL}^{min} = F_{SS}(1 - \epsilon, \alpha_A).
$$

Here, $1 - \epsilon$ denotes the required minimum probability for a $\mathcal{SS}$, and $\alpha_A$ denotes the fraction of network resources controlled by the adversary. Note that $F_{\mathcal{SS}}$ is a useful function to ensure Faithful Fault Tolerance Condition with desired probability guarantee.

\begin{proposition} \label{proposition1}
A \ourapproach\ protocol is in a secure state for a given $\alpha_A$ with probability of at-least $1 - \epsilon$ if the minimum number of nodes in $CSL$ is
$$
n_{CSL}^{min} = F_{SS}(1 - \epsilon, \alpha_A),
$$
where
$$
F_{SS} = argmin_{n_{CSL}} \{(1 - F(n_f; n_{CSL},\alpha_A))\leq \epsilon\}.
$$
Here $F(n_f; n_{CSL}, \alpha_A)$ is an appropriately defined function in the proof and gives the probability of $n_A < n_f$.
\end{proposition}




\begin{proof}
The probability that a participant mines a block in $ACL$ is directly proportional to its fraction of computational power in the system, i.e., 
$$
\Pr(\mbox{Mine a block in ACL}) \propto \alpha_{participant}.
$$
This follows by observing that the size of the committee, i.e., $n_{CSL}$ is finite. Let $X$ be the discrete random variable denoting the number of validators controlled by the adversary. Consider the following standard binomial distribution for deriving the probability of the number of validators controlled by the adversary $A$ as,
$$
    F(k; n_{CSL}, \alpha_A) = \Pr(k; n_{CSL}, \alpha_A) = \Pr(X = k) \nonumber
$$
$$
\mbox{s.t.~}\Pr(X = k) = (^{n_{CSL}}_k)(\alpha_A)^k(1 - \alpha_A)^{(n_{CSL} - k)}
$$

Now, let $F$ be the corresponding CDF defined as,
$$
F(x;n_{CSL}, \alpha_A) = \Pr(X \leq x).
$$
$F(n_f;n_{CSL}, \alpha_A)$ gives the probability that the system is in a secure state, i.e. $n_A < n_f $, which depends on the value of $n_{CSL}$ and $\alpha_A$ from the binary distribution.

\noindent With this we set
$$
F_{SS}(1-\epsilon,\alpha_A) = argmin_{n_{CSL}} \{(1- F(n_f ; n_{CSL},\alpha_A))\leq \epsilon\}.
$$
\end{proof}

\smallskip

\smallskip
\noindent\textbf{Selecting Safe Committee Size.} \label{selecting:n}
$\epsilon$ denotes the probability of a system not in a secure state, i.e. $\epsilon = \Pr(\Pi_{\mathcal{TM}} \mbox{ not in }  \mathcal{SS})$. Observe that the probability of a node being added to $CSL$ is directly proportional to $\alpha_{participant}$, i.e., the resources it holds in the system. Since, each node in the $CSL$ can either belong to the rational participants with probability $1 - \alpha_A$ or to a byzantine adversary with probability $\alpha_A$, we can model this distribution as a binomial distribution. We know that, from Proposition~\ref{proposition1}, the protocol will not be in a secure state if the number of nodes controlled by the adversary in the committee are equal to or more than $n_f$, i.e. $\alpha_A \geq n_f$. 
Thus, $\epsilon$ becomes the binomial distribution of $n_A$ successes in $n_{CSL}$ independent experiments, with $\frac{\alpha_A}{n_{CSL}}$ as the probability of a each success. Based on different values of $\alpha_A$ and probability guarantees of secure state, we obtain the values of $n_{CSL}^{min}$\footnote{For additional details, we refer the reader to Proposition~\ref{proposition1}}. We consider this analysis for probability guarantees, for system being in a secure state, ranging from $2 \times 10^{-4}$ to $2 \times 10^{-6}$. Note that in Bitcoin, for 10\% computational power with adversary, the system is said to be secure with probability of $2 \times 10^{-4}$ i.e., 1 in 5000. We consider the values of $\alpha_A$, the resources controlled by the adversary, in the range 1\% to 18\%. This range follows from observation that the most computational resources owned by a single mining pool in Bitcoin is 17\%.


Following from this, observe that this upper bound on $\alpha_A$, becomes closer to $\frac{n_f}{n_{CSL}}$ as we increase $n_{CSL}$. In general, for any Byzantine agreement protocol to function correctly it requires, at the least, that the number of participants controlled by the adversary $n_A$, be less than one-third of total nodes, $n_{CSL}$. For instance, $n_{f} = \frac{n_{CSL} + 2}{3}$ for PBFT~\cite{pbft}. Therefore, our upper bound for the fraction of network-resources controlled by an Adversary A is around $\frac{1}{3}$. In most blockchain protocols, using consensus mechanisms like Proof-of-Work, Proof-of-Stake, etc., this bound is $\frac{1}{2}$ (Bitcoin). Although, one should note that in practical scenarios, $\alpha_A$ does not reach over $\frac{1}{5}$ or 20\%. For reference, the most computational resources owned by any miner in Bitcoin is 17\%.
Since our consideration is much higher than this percentage, we can say that the bounds in \ourapproach\ are reasonable for practical scenarios.

\subsection{Discussion}

We highlight that a \ourapproach\ protocol can be designed using any Proof-of-X consensus protocol in the first layer, inheriting its security. Similarly, any BFT protocol in the second layer can be used. In our analysis of \ourapproach, considering three types of nodes, we proved that the protocol will be Nash-Incentive-Compatible (Definition~\ref{def:ics}) for three certain conditions according to Theorem \ref{thm::PSNE}. We have also proved that any \ourapproach\ protocol is in a secure state (Definition~\ref{def:SS}) with high probability under reasonable assumptions and pre-defined conditions on protocol parameters. 


One must note that PeerCensus \cite{peercensus}, Solida \cite{solida}, Hybrid Consensus \cite{hybrid-consensus}, and we believe, a few other protocols similar to Algorand \cite{algorand} are a special instantiation of \ourapproach. In these special cases, we say that PeerCensus and Solida use a Proof-of-Work based protocol whereas Hybrid Consensus and Algorand use a Proof-of-Stake protocol in the $ACL$. However, these protocols will need formal analysis to prove so and this is left for future work. We believe our framework has further potential to be more universal and include major protocols as its instantiations.
\section{ASHWAChain}
ASHWAChain~\cite{Arora2020} is an example of as multi-layered blockchain protocol that fits in the Tiramisu framework. \ourproto\ is a committee-based blockchain protocol in which the committee is being updated regularly. We present a formal analysis of the protocol in the full version of our paper~\cite{tiramisu}. The Access Control Layer of \ourproto\ use Proof-of-Work as a method of sybil-resistance and PBFT~\cite{pbft} towards achieving byzantine agreement in the Consensus Layer. As per analysis using the Tiramisu framework, we show that after selecting secure protocol parameters that ensure game-theoretic soundness the protocol can achieve a throughput of more than 700 transactions per second, more than the claim in the original paper by selecting a smaller committee size but greater Proof-of-Work difficulty. We remark that analysis with out framework is also useful in finding optimal parameters for multi-layered blockchain protocols. Figures~\ref{fig:spvspop} and~\ref{fig:tps} show the effect of committee size on security and performance for ASHWAChain.

\section{Conclusion}\label{sec:conclucion}
In this work, we formalize a layered approach towards blockchain consensus, \ourapproach. The layered approach allows us to combine permissioned and permissionless blockchain protocols to design protocols that provide both scalability and decentralization (Section~\ref{sec:tiramisu}). We use separate layers of consensus protocols, much like a Tiramisu. Along with a security analysis, we use a game-theoretic analysis in the BAR model to show that our framework assures security in the presence of Byzantine adversary with limited network resources while providing a better throughput than traditional permissionless blockchain protocols (Section ~\ref{sec:tiramisu_analysis}). Lastly, we proved that under certain assumptions for the protocol parameters, there exists a Pure Strategy Nash Equilibrium in which the rational players validate the transactions before signing any block. We believe this abstract approach of Tiramisu is useful for blockchain practitioners to securely design layered blockchain protocols for greater performance. 
\bibliographystyle{unsrtnat}
\bibliography{references}  





\section{ASHWAChain Details}

\begin{algorithm}[t]
\DontPrintSemicolon
\Begin{
    \eIf{p = Primary}{
        Creates any operation and broadcasts\;
    }{
        Validate operation, sign and broadcast\;
    }
    wait(\mbox{Latency Time})\;

    \eIf{signed by super-majority}{
        Append $operation$ to $O$\;
        Commit $operation$ according to its commit rule\;
    }{
        Nothing is committed, RESET\;
    }
}
\caption{\ourproto's agreement process for identity $p$ \label{algo:agreement}}
\end{algorithm}

\begin{algorithm}[!t]
\DontPrintSemicolon
\textbf{Validate} \texttt{powBlock}$(b)$:
\Begin{
    $b^* \leftarrow \mbox{latest block in } PowChain$\;
    \eIf{$b$ is child of $b^*$ and $b$ is valid}{ 
        return Valid\;
    }{
        return Invalid\;
    }
}

\textbf{Validate} \texttt{comBlock}$(c)$:
\Begin{
    \eIf{All $n_{CSL}$ nodes in $c$ match\\ with the \textsc{PowChain} and $c$ is valid}{ 
        return Valid\;
    }{
        return Invalid\;
    }
}
\textbf{Validate} \texttt{txBlock}$(t)$:
\Begin{
    \eIf{$signatures$ and all $tx$s are valid}{ 
        return Valid\;
    }{
        return Invalid\;
    }
}
\textbf{Commit} \texttt{powBlock}$(b)$:
\Begin{
    Append \texttt{powBlock}$(b)$ in $O$\;
    Append $b$ to \textsc{PowChain}\;
}
\textbf{Commit} \texttt{comBlock}(c):
\Begin{
    Append \texttt{comBlock}$(c)$ in $O$\;
    Append $c$ to \textsc{ComChain}\;
}
\textbf{Commit} \texttt{txBlock}$(t)$:
\Begin{
    Append \texttt{txBlock}$(t)$ to $O$\;
    \For{$tx \in t$}{
        $B(from) \leftarrow B(from) - coins$\;
        $B(to) \leftarrow B(to) + coins$\;
    }
    Append the list of validators of $t$ in $T$\; 
    \For{$v \in T(t)$; $txFee \leftarrow \mbox{Total transaction fee}$;\\ $nSig \leftarrow \mbox{Total signatures}$}{
        $B(v) \leftarrow B(v) + \frac{txFee + \mbox{Block Reward}}{nSig}$\;
    }
    Append $t$ to \textsc{TxChain}
}

\caption{\ourproto: Predefined operations to modify the shared state in $CSL$.}\label{algo:operations}
\end{algorithm}

\end{document}